\documentclass[aps,onecolumn,showpacs,showkeys,nofootinbib]{revtex4}
\usepackage{epsfig}
\usepackage{amsmath}
\usepackage{amsfonts}
\usepackage{amssymb}
\usepackage{graphicx}
\usepackage{colordvi}
\begin{document}

\title{Weak Gravitational Lensing in Fourth Order Gravity}

\author{$^1$A. Stabile\footnote{arturo.stabile@gmail.com}, An. Stabile\footnote{anstabile@gmail.com}}

\affiliation{$^1$Dipartimento di Ingegneria, Universita' del
Sannio, Palazzo Dell'Aquila Bosco Lucarelli, Corso Garibaldi, 107
- 82100, Benevento, Italy}

\begin{abstract}

For a general class of analytic $f(R,R_{\alpha\beta}R^{\alpha\beta},R_{\alpha\beta\gamma\delta}R^{\alpha\beta\gamma\delta})$ we discuss the gravitational lensing in the Newtonian Limit of theory. From the properties of Gauss Bonnet invariant it is successful  to consider only one curvature invariants between the Ricci and Riemann tensor. Then we analyze the dynamics of photon embedded in a gravitational field of a generic $f(R,R_{\alpha\beta}R^{\alpha\beta})$-Gravity. The metric is time independent and spherically symmetric. The metric potentials are Schwarzschild-like, but there are two additional Yukawa terms linked to derivatives of $f$ with respect to two curvature invariants. Considering first the case of a point-like lens, and after the one of a generic matter distribution of lens, we study the deflection angle and the angular position of images. Though the additional Yukawa terms in the gravitational potential modifies dynamics with respect to the General Relativity, the geodesic trajectory of photon is unaffected by the modification if we consider only $f(R)$-Gravity. While we find different results (deflection angle smaller than one of General Relativity) only thank to introduction of a generic function of Ricci tensor square. Finally we can affirm the lensing phenomena for all $f(R)$-Gravities are equal to the ones known of General Relativity. We conclude the paper showing and comparing the deflection angle and position of images for $f(R,R_{\alpha\beta}R^{\alpha\beta})$-Gravity with respect to the ones of General Relativity.

\end{abstract}
\maketitle

\section{Introduction}

Despite of all nice results of General Relativity (GR),
the study of possible modifications of Einstein's theory
of Gravity has a long history which reaches back to the
early 1920s \cite{weyl, weyl1, pauli, bach, eddington,
lanczos}. While the proposed early amendments of Einstein's
theory were aimed toward the unification of Gravity with
the other interactions of Physics, the recent interest in
such modifications comes from cosmological observations
(for a comprehensive review, see \cite{schmidt}). In fact
the presence of the Big Bang singularity, the flatness and
horizon problems \cite{guth} led to the statement that
Cosmological Standard Model, based on GR and Standard Model of
Particle Physics, is inadequate to describe the Universe at
extreme regimes. Besides from Quantun Field Theory point view, GR is a
\emph{classical} theory which does not work as a fundamental
theory, when one wants to achieve a full quantum description of
spacetime (and then of gravity).

The astrophysical and cosmological observations usually lead to the
introduction of additional \emph{ad-hoc} concepts like Dark
Energy/Matter if interpreted within Einstein's theory. In fact the principal
physical aspects are the cosmic acceleration and the flat galactic rotation curves.
These aspects could be interpreted as a first signal of a breakdown of GR at astrophysical and
cosmological scales \cite{riess, ast, clo, ber, spe, anderson1, anderson2, cap-card-tro2, CCT} and led
to the proposal of several alternative modifications of the
underlying gravity theory (see \cite{noj-odi6} for the review).

While it is very natural (from the theoretical point of view) to extend Einstein's Gravity to theories
with additional geometric degrees of freedom, (see for example
\cite{heh-von-ker-nes, heh-mcc-mie-nee, trautman} for general
surveys on this subject as well as \cite{puetzfeld} for a list of
works in a cosmological context), recent attempts focused on the
old idea of modifying the gravitational Lagrangian in a purely
metric framework, leading to higher order field equations. As such an
approach is the so-called Extended Theories of Gravity
(ETG) which have become a sort of paradigm in the study of
gravitational interaction. They are based on corrections and
enlargements of the Einstein theory. The paradigm consists,
essentially, in adding higher order curvature invariants and
minimally or non-minimally coupled scalar fields into dynamics
which come out from the effective action of quantum gravity
\cite{buc-odi-sha}. A sub class of ETG are the Fourth Order Gravities (FOG) where
we do not consider further scalar fields but only the curvature invariants.

The motivation to modify the GR come from the issue of a full
recovering of the Mach principle which leads to assume a varying
gravitational coupling. The principle states that the local
inertial frame is determined by some average of the motion of
distant astronomical objects \cite{bondi}. This fact implies that
the gravitational coupling can be scale-dependent and related to
some scalar field. As a consequence, the concept of ``inertia''
and the Equivalence Principle have to be revised. For example, the
Brans-Dicke theory \cite{bra-dic} is a serious attempt to define
an alternative theory to the Einstein gravity: it takes into
account a variable Newton gravitational coupling, whose dynamics
is governed by a scalar field non-minimally coupled to the
geometry. In such a way, Mach's principle is better implemented
\cite{bra-dic, cap-der-rub-scu, sciama}. As already mentioned, corrections to the gravitational Lagrangian
were already studied by several authors \cite{weyl1, eddington, lanczos}
soon after the GR was proposed. Developments in the 1960s and 1970s \cite{buchdahl1,
dewitt, bicknell, havas, stel}, partially motivated by the
quantization schemes proposed at that time, made clear that
theories containing {\it only} a Ricci scalar square term in the Lagrangian were
not viable with respect to their weak field behavior. Another concern which comes with generic \emph{higher order
gravity} (HOG) theories is linked to the initial value problem. It
is unclear if the prolongation of standard methods can be used in
order to tackle this problem in every theory. Hence it is doubtful
that the Cauchy problem could be properly addressed in the near
future, for example within the theories with inverse of Ricci scalar, if one takes into
account the results already obtained in fourth order theories
stemming from a quadratic Lagrangian \cite{tey-tou, dur-ker}.

Besides, every unification scheme as Superstrings, Supergravity or
Grand Unified Theories, takes into account effective actions where
non-minimal couplings to the geometry or higher order terms in the
curvature invariants are present. Such contributions are due to
one-loop or higher loop corrections in the high curvature regimes
near the full (not yet available) quantum gravity regime
\cite{buc-odi-sha}. Specifically, this scheme was adopted in order
to deal with the quantization on curved spacetimes and the result
was that the interactions among quantum scalar fields and
background geometry or the gravitational self-interactions yield
corrective terms in the Hilbert-Einstein Lagrangian
\cite{bir-dav}.

From a conceptual viewpoint, there are no \emph{a priori} reasons to
restrict the gravitational Lagrangian to
the linear function of the Ricci scalar, minimally coupled with
matter \cite{mag-fer-fra}. Since all curvature invariants are at
least second order differential, the corrective terms in the field
equations will be always at least fourth order. That is why generally we call them
higher order terms (with respect to the terms of GR).

The idea to extend Einstein's theory of gravitation is fruitful
and economic also with respect to several attempts which try to
solve problems by adding new and, most of times, unjustified
ingredients in order to give a self-consistent picture of
dynamics. The today observed accelerated expansion of the Hubble
flow and the missing matter of astrophysical large scale
structures, are primarily enclosed in these considerations. Both
the issues could be solved changing the gravitational sector,
\emph{i.e.} the \emph{l.h.s.} of field equations. The philosophy
is alternative to add new cosmic fluids (new components in the
\emph{r.h.s.} of field equations) which should give rise to
clustered structures (dark matter) or to accelerated dynamics
(dark energy) thanks to exotic equations of state. In particular,
relaxing the hypothesis that gravitational Lagrangian has to be a
linear function of the Ricci curvature scalar, like in the
Hilbert-Einstein formulation, one can take into account an
effective action where the gravitational Lagrangian includes other
scalar invariants. In summary, the general features of ETGs are that the Einstein
field equations result to be modified in two senses: $i)$ geometry
can be non-minimally coupled to some scalar field, and / or $ii)$
higher than second order derivative terms in the metric come out.
In the former case, we generically deal with scalar-tensor
theories of gravity; in the latter, we deal with higher order
theories. However combinations of non-minimally coupled and
higher-order terms can emerge as contributions in effective
Lagrangians. In this case, we deal with higher-order-scalar-tensor
theories of Gravity.

Due to the increased complexity of the field equations in this
framework, the main amount of works dealt with some formally
equivalent theories, in which a reduction of the order of the
field equations was achieved by considering the metric and the
connection as independent fields \cite{ama-elg-mot-mul,
mag-fer-fra, all-bor-fra, sot1, sot-lib}. In addition, many
authors exploited the formal relationship to scalar-tensor
theories to make some statements about the weak field regime,
which was already worked out for scalar-tensor theories more than
ten years ago \cite{dam-esp}. Moreover other authors exploited a systematic analysis of such theories were performed at short scale and in the low energy limit \cite{olmo1, olmo2, olmo3, Damour:Esposito-Farese:1992, clifton, odintsov, PRD, PRD1, PRD2, Stabile_Capozziello, CCCT}. In particular the FOG has been studied in the so-called Newtonian Limit (case of weak field and small velocity) and in the Weak Field Limit (case of Gravitational waves) \cite{minko}. In the first case we found a modification of gravitational potential, while in the second one the massive propagation of waves. After these preliminary studies it needs to check the new theories by applying them in the realistic model. Then the galactic rotation curve \cite{CCT, BHL, BHL1, Stabile_scelza} or the motion of body in the Solar System have been evaluated.

It is worth remembering that one of the first experimental
confirmations of Einsteinian theory of Gravity was
the deflection of light. Since then, gravitational lensing (GL) has become one of
the most useful successes of GR and
it represents nowadays a powerful tool \cite{lensing}. In this paper, we investigate the equation for the photon deflection considering the Newtonian Limit of a general
class of $f(X,Y,Z)$-Gravity where $f$ is a unspecific function of $X\,=\,R$ (Ricci scalar), $Y\,=\,R_{\alpha\beta}R^{\alpha\beta}$ (Ricci tensor square) and $Z\,=\,R_{\alpha\beta\gamma\delta}R^{\alpha\beta\gamma\delta}$ (Rieman tensor square).

The starting point of paper is the metric tensor solution of modified Einstein equation \cite{PRD2} and we analyze the dynamics of photon. Particularly in Section II we introduce the $f(X,Y,Z)$-Gravity and its field equations resuming shortly the Newtonian limit approach, while the Section III-A is devoted to study of photon traveling embedded in a gravitational field. In this section we approach first the deflection angle by a point-like source and after we reformulate the lensing problem by a generic matter distribution (Section III-B). Moreover we rewrite the equation lens (Section III-C) and find the corrections to the position of images in the case of point-like lens. Finally in Section IV there are comments and conclusions.

\section{The Newtonian Limit of Fourth Order Gravity}

Let us start with a general class of FOG given
by the action

\begin{eqnarray}\label{HOGaction}\mathcal{A}=\int d^{4}x\sqrt{-g}\biggl[f(X,Y,Z)+\mathcal{X}\mathcal{L}_m\biggr]
\end{eqnarray}
where $f$ is an unspecified function of curvature invariants. The term $\mathcal{L}_m$ is the minimally coupled ordinary matter contribution. In the metric approach, the field equations are obtained by varying (\ref{HOGaction}) with respect to $g_{\mu\nu}$. We get

\begin{eqnarray}\label{fieldequationFOG}
f_XR_{\mu\nu}-\frac{f}{2}g_{\mu\nu}-f_{X;\mu\nu}+g_{\mu\nu}\Box
f_X+2f_Y{R_\mu}^\alpha
R_{\alpha\nu}-2[f_Y{R^\alpha}_{(\mu}]_{;\nu)\alpha}+\Box[f_YR_{\mu\nu}]+[f_YR_{\alpha\beta}]^{;\alpha\beta}g_{\mu\nu}&&
\nonumber\\\nonumber\\+2f_ZR_{\mu\alpha
\beta\gamma}{R_{\nu}}^{\alpha\beta\gamma}-4[f_Z{{R_\mu}^{\alpha\beta}}_\nu]_{;\alpha\beta}&&\,=\,
\mathcal{X}\,T_{\mu\nu}
\end{eqnarray}
where
$T_{\mu\nu}\,=\,-\frac{1}{\sqrt{-g}}\frac{\delta(\sqrt{-g}\mathcal{L}_m)}{\delta
g^{\mu\nu}}$ is the energy-momentum tensor of matter,
$f_X\,=\,\frac{df}{dX}$, $f_Y\,=\,\frac{df}{dY}$,
$f_Z\,=\,\frac{df}{dZ}$, $\Box={{}_{;\sigma}}^{;\sigma}$ and
$\mathcal{X}\,=\,8\pi G$\footnote{Here we use the convention
$c\,=\,1$.}. The convention for Ricci's tensor is
$R_{\mu\nu}={R^\sigma}_{\mu\sigma\nu}$, while for the Riemann
tensor is
${R^\alpha}_{\beta\mu\nu}=\Gamma^\alpha_{\beta\nu,\mu}+...$. The
affinities are the usual Christoffel symbols of the metric:
$\Gamma^\mu_{\alpha\beta}=\frac{1}{2}g^{\mu\sigma}(g_{\alpha\sigma,\beta}+g_{\beta\sigma,\alpha}
-g_{\alpha\beta,\sigma})$. The adopted signature is $(+---)$ (for details, see \cite{landau}).

The paradigm of the Newtonian limit starts from the development of the
metric tensor (and of all additional quantities in the theory)
with respect to the dimensionless quantity $v$ but considering only
first term of $tt$- and $ij$-component of metric tensor
$g_{\mu\nu}$ (for details, see \cite{PRD1}). The
develop of the metric is the following

\begin{eqnarray}\label{me0}
{ds}^2\,=\,(1+2\Phi)dt^2-(1-2\Psi)\delta_{ij}dx^idx^j
\end{eqnarray}
where $\Phi$ and $\Psi$ are proportional to $v^2$. The set of coordinates\footnote{The Greek index runs from $0$ to $3$; the Latin index runs from $1$ to $3$.} adopted is
$x^\mu\,=\,(t,x^1,x^2,x^3)$. The curvature invariants $X$, $Y$,
$Z$ become

\begin{eqnarray}
\left\{\begin{array}{ll}
X\,\sim\,X^{(2)}+\dots\\\\
Y\,\sim\,Y^{(4)}+\dots\\\\
Z\,\sim\,Z^{(4)}+\dots
\end{array}\right.
\end{eqnarray}
and the function $f$ and its partial derivatives ($f_X$, $f_{XX}$, $f_{Y}$ and $f_{Z}$) can be substituted by their corresponding Taylor develop. In the case of $f$ we have

\begin{eqnarray}
f(X,Y,Z)\,&\sim&\,f(0)+f_X(0)X^{(2)}+\frac{1}{2}f_{XX}(0){X^{(2)}}^2+f_X(0)X^{(4)}+f_Y(0)Y^{(4)}+f_Z(0)Z^{(4)}+\dots
\end{eqnarray}
and analogous relations for derivatives are
obtained.

From the lowest order of field equations (\ref{fieldequationFOG}) we have

\begin{eqnarray}\label{PPN-field-equation-general-theory-fR-O0}
f(0)\,=\,0
\end{eqnarray}
while in the Newtonian Limit ($\propto\,v^2$) we have\footnote{Throughout the paper we assume always $f_X(0)\,>\,0$, and therefore we may set $f_X(0)\,=\,1$ without loss of generality.}

\begin{eqnarray}\label{NL-field-equation_2}
\left\{\begin{array}{ll}
(\triangle-{m_2}^2)\triangle\Phi+\biggl[{m_2}^2-\frac{{m_1}^2+2{m_2}^2}{3{m_1}^2}\triangle\biggr]
X^{(2)}\,=\,-2\,{m_2}^2\mathcal{X}\,\rho\\\\
\triangle\Psi\,=\,\int
d^3\mathbf{x}'\mathcal{G}_2(\mathbf{x},\mathbf{x}')\biggl(\frac{{m_2}^2}{2}-\frac{{m_1}^2+2{m_2}^2}{6{m_1}^2}
\triangle_{\mathbf{x}'}\biggr)X^{(2)}(\mathbf{x}')\\\\
(\triangle-{m_1}^2)X^{(2)}\,=\,{m_1}^2\mathcal{X}\,\rho
\end{array}\right.
\end{eqnarray}
where $X^{(2)}$ is the Ricci scalar at Newtonian order, $\rho$ is the matter density and $\mathcal{G}_2$ is the Green function of field operator $\triangle-{m_2}^2$. The quantities ${m_i}^2$ are linked to derivatives of $f$ with respect
to the curvature invariants $X$, $Y$ and $Z$

\begin{eqnarray}\label{mass_definition}
\left\{\begin{array}{ll}
{m_1}^2\,\doteq\,-\frac{1}{3f_{XX}(0)+2f_Y(0)+2f_Z(0)}\\\\
{m_2}^2\,\doteq\,\frac{1}{f_Y(0)+4f_Z(0)}
\end{array}\right.
\end{eqnarray}

By solving the field equations (\ref{NL-field-equation_2}), if ${m_i}^2\,>\,0$ for $i\,=\,1\,,2$, the proper time interval, generated by a
point-like source with mass $M$, is (for details, see \cite{PRD1, PRD2})

\begin{eqnarray}\label{me1}
{ds}^2\,=\,\biggl[1-r_g\biggl(\frac{1}{|\textbf{x}|}
+\frac{1}{3}\frac{e^{-\mu_1|\mathbf{x}|}}{|\mathbf{x}|}
-\frac{4}{3}\frac{e^{-\mu_2|\mathbf{x}|}}{|\mathbf{x}|}\biggr)
\biggr]dt^2-
\biggl[1+r_g\biggl(\frac{1}{|\textbf{x}|}
-\frac{1}{3}\frac{e^{-\mu_1|\mathbf{x}|}}{|\mathbf{x}|}
-\frac{2}{3}\frac{e^{-\mu_2|\mathbf{x}|}}{|\mathbf{x}|}\biggr)\biggr]\delta_{ij}dx^idx^j
\end{eqnarray}
where $r_g\,=\,2GM$ is the Schwarzschild radius and
$\mu_i\,\doteq\,\sqrt{|{m_i}^2|}$. The field equations (\ref{NL-field-equation_2}) are valid
for any values of quantities ${m_i}^2$, while the Green functions of field operator $\triangle-{m_i}^2$ admit two different behaviors if ${m_i}^2\,>\,0$ or ${m_i}^2\,<\,0$. The possible choices of Green function, for spherically symmetric systems (\emph{i.e.} $\mathcal{G}_i(\mathbf{x},\mathbf{x}')\,=\,\mathcal{G}_i(|\mathbf{x}-\mathbf{x}'|)$), are the following

\begin{eqnarray}\label{green_function}
\mathcal{G}_i(\mathbf{x},\mathbf{x}')\,=\,\left\{\begin{array}{ll}-\frac{1}{4\pi}\frac{e^{-\mu_i|\mathbf{x}-\mathbf{x}'|}}
{|\mathbf{x}-\mathbf{x}'|}\,\,\,\,\,\,\,\,\,\,\,\,\,\,\,\,\,\,\,\,\,\,\,\,\,\,\,\,\,\,\,\,\,\,\,\,\,\,\,\,\,\,\,\,\,
\,\,\,\text{if}\,\,\,\,\,\,\,\,\,\,\,\,\,\,{m_i}^2\,\,>\,0
\\\\
-\frac{1}{4\pi}\frac{\cos
\mu_i|\mathbf{x}-\mathbf{x}'|+ \sin
\mu_i|\mathbf{x}-\mathbf{x}'|}{|\mathbf{x}-\mathbf{x}'|}
\,\,\,\,\,\,\,\,\,\,\,\,\,\,\,\text{if}\,\,\,\,\,\,\,\,\,\,\,\,\,\,{m_i}^2\,<\,0\end{array}\right.
\end{eqnarray}
The first choice in (\ref{green_function}) corresponds to Yukawa-like behavior, while the second one to the oscillating case. Both expressions are a generalization of the usual gravitational potential ($\propto |\mathbf{x}|^{-1}$), and when ${m_i}^2\,\rightarrow\,\infty$ (\emph{i.e.} $f_{XX}(0)\,,f_Y(0)\,,f_Z(0)\,\rightarrow\,0$ from the (\ref{mass_definition})) we recover the field equations of GR. Independently of algebraic sign of ${m_i}^2$ we can introduce two scale lengths ${\mu_i}^{-1}$. We note that in the case of $f(X)$-Gravity we obtained only one scale length (${\mu_1}^{-1}$ with $f_Y(0)\,=\,f_Z(0)\,=\,0$) on the which the Ricci scalar evolves \cite{PRD1, PRD2}, but in $f(X,Y,Z)$-Gravity we have an additional scale length ${\mu_2}^{-1}$ on the which the Ricci tensor evolves.

Often for spherically symmetric problems it is convenient
rewriting the metric (\ref{me1}) in the so-called standard coordinates system\footnote{Generally the set of coordinates $(t,r,\theta,\phi)$ are called standard coordinates if the metric is expressed as ${ds}^2\,=\,g_{tt}(t,r)\,dt^2+g_{rr}(t,r){dr}^2-r^2d\Omega$ while if one has ${ds}^2\,=\,g_{tt}(t,\mathbf{x})\,dt^2+g_{ij}(t,\mathbf{x})dx^idx^j$ (like the solution (\ref{me1})) the set $(t,x^1,x^2,x^3)$ is called isotropic coordinates \cite{weinberg}.} (the
usual form in which we write the Schwarzschild solution). By
introducing a new radial coordinate
$\tilde{r}\,=\,|\tilde{\mathbf{x}}|$ as follows

\begin{eqnarray}\label{condit_transf}
\biggl[1-2\Psi(r)\biggr]r^2\,=\,\tilde{r}^2
\end{eqnarray}
the relativistic invariant (\ref{me1}) becomes\footnote{The metrics (\ref{standard_metric}) and (\ref{me1}) represent the same space-time at first order of $r_g/r$.}

\begin{eqnarray}\label{standard_metric}
  ds^2\,=\,\biggl[1-\frac{r_g}{r}\biggl(1+\frac{1}{3}\,e^{-\mu_1r}-\frac{4}{3}\,e^{-\mu_2r}\biggr)\biggr]dt^2-\biggl[1+\frac{r_g}{r}
  \biggl(1-\frac{\mu_1r+1}{3}\,e^{-\mu_1r}
  -\frac{2(\mu_2r+1)}{3}\,e^{-\mu_2r}\biggr)\biggr]dr^2-r^2d\Omega
\end{eqnarray}
where $d\Omega\,=\,d\theta^2+\sin^2\theta\,d\phi^2$ is the solid
angle and renamed the radial coordinate $\tilde{r}$.

\section{The Gravitational Lensing by $f(X,Y,Z)$-Gravity}

\subsection{Point-Like Source}

The Lagrangian of photon in the gravitational field with metric
(\ref{standard_metric}) is

\begin{eqnarray}\label{phot_lagr}
  \mathcal{L}\,=\,\frac{1}{2}\biggl[\biggl(1-\frac{r_g}{r}\,\Xi(r)\biggr)
  \dot{t}^2-\biggl(1+\frac{r_g}{r}\,\Lambda(r)\biggr)\dot{r}^2
  -r^2\dot{\theta}^2-r^2\sin^2\theta\dot{\phi}^2\biggr]
\end{eqnarray}
where
$\Xi(r)\,\doteq\,1+\hat{\Xi}(r)\,\doteq\,1+\frac{1}{3}\,e^{-\mu_1r}-\frac{4}{3}\,e^{-\mu_2r}$,
$\Lambda(r)\,\doteq\,1+\hat{\Lambda}(r)\,\doteq\,1-\frac{\mu_1r+1}{3}\,e^{-\mu_1r}-\frac{2(\mu_2r+1)}{3}\,e^{-\mu_2r}$
and the dot represents the derivatives with respect to the affine
parameter $\lambda$. Since the variable $\theta$ does not have dynamics
($\ddot{\theta}\,=\,0$) we can choose for simplicity
$\theta\,=\,\pi/2$. By applying the Euler-Lagrangian equation to
Lagrangian (\ref{phot_lagr}) for the cyclic variables $t$, $\phi$
we find two motion constants

\begin{eqnarray}\label{motion_constant}
\left\{\begin{array}{ll}
\frac{\partial\mathcal{L}}{\partial\dot{t}}\,=\,\biggl(1-\frac{r_g}{r}\,\Xi(r)\biggr)\dot{t}\,\doteq\,\mathcal{T}
\\\\
\frac{\partial\mathcal{L}}{\partial\dot{\phi}}\,=\,-r^2\dot{\phi}\,\doteq\,-J
\end{array}\right.
\end{eqnarray}
and respect to $\lambda$ we find the "energy" of
Lagrangian\footnote{The (\ref{phot_lagr}) is a quadratic form, so
it corresponds to its Hamiltonian.}

\begin{eqnarray}\label{energy}
  \mathcal{L}\,=\,0
\end{eqnarray}
By inserting the equations (\ref{motion_constant}) into (\ref{energy}) we find a differential equation for $\dot{r}$

\begin{eqnarray}\label{diffequ}
  \dot{r}_{\pm}\,=\,\pm\,\mathcal{T}\,\sqrt{\frac{1}{1+\frac{r_g}{r}\,\Lambda(r)}\biggl[\frac{1}{1-\frac{r_g}{r}\,\Xi(r)}-
  \frac{J^2}{r^2}\bigg]}
\end{eqnarray}
$\dot{r}_+$ is the solution for leaving photon, while $\dot{r}_-$
is one for incoming photon. Let $r_0$ be a minimal distance from
the lens center (Fig. \ref{deflection}). We must impose the condition
$\dot{r}_{\pm}(r_0)\,=\,0$ from the which we find

\begin{eqnarray}\label{Jcond}
  J^2\,=\,\frac{{r_0}^2\mathcal{T}^2}{1-\frac{r_g}{r_0}\,\Xi(r_0)}
\end{eqnarray}
Now the deflection angle $\alpha$ (Fig. \ref{deflection}) is defined by following relation

\begin{eqnarray}\label{angle_deflec}
  \alpha\,=\,&&-\pi+\phi_{fin}\,=\,-\pi+\int_0^{\phi_{fin}}d\phi\,=\,-\pi+\int_{\lambda_{in}}^{\lambda_{fin}}
  \dot{\phi}\,d\lambda\,=\,-\pi+
  \int_{\lambda_{in}}^{\lambda_0}\dot{\phi}\,d\lambda+\int_{\lambda_0}^{\lambda_{fin}}\dot{\phi}\,d\lambda\,=\nonumber\\
  \nonumber\\
  &&-\pi+\int_{\infty}^{r_0}\frac{\dot{\phi}}{\dot{r}_-}\,dr+\int_{r_0}^{\infty}\frac{\dot{\phi}}{\dot{r}_+}\,dr\,=\,
  -\pi+2\int_{r_0}^{\infty}\frac{\dot{\phi}}{\dot{r}_+}\,dr
\end{eqnarray}
where $\lambda_0$ is the value of $\lambda$ corresponding to the minimal value ($r_0$) of radial coordinate $r$. By putting the expressions of $J$, $\dot{\phi}$ and $\dot{r}_+$ into (\ref{angle_deflec}) we get the deflection angle

\begin{eqnarray}\label{angle_deflec_1}
  \alpha\,=\,-\pi+2\int_{r_0}^\infty\frac{dr}{r\sqrt{\frac{1}{1+\frac{r_g}{r}\,\Lambda(r)}\biggl[\frac{1-\frac{r_g}
  {r}\,\Xi(r_0)}{1-\frac{r_g}{r}\,\Xi(r)}\frac{r^2}{{r_0}^2}-1\bigg]}}
\end{eqnarray}
which in the case $r_g/r\,\ll\,1$\footnote{We do not consider the GL generated by a black hole.} becomes

\begin{eqnarray}\label{angle_deflec_2}
  \alpha\,=\,2\,r_g\biggl[\frac{1}{r_0}+\mathcal{F}_{\mu_1,\,\mu_2}(r_0)\biggr]
\end{eqnarray}
where

\begin{eqnarray}\label{fun1}
  \mathcal{F}_{\mu_1,\,\mu_2}(r_0)\,\doteq\,\frac{1}{2}\int_{r_0}^\infty
  \frac{r_0r^2[\hat{\Lambda}(r)-\hat{\Xi}(r)]+r^3\hat{\Xi}(r_0)-{r_0}^3\hat{\Lambda}(r)}{r^3(r^2-{r_0}^2)
  \sqrt{1-\frac{{r_0}^2}{r^2}}}\,dr
\end{eqnarray}
From the definition of $\hat{\Xi}$ and $\hat{\Lambda}$ we note that in the case $f(X,Y,Z)\,\rightarrow\,X$ we obtain $\mathcal{F}_{\mu_1,\,\mu_2}(r_0)\,\rightarrow\,0$. In a such way we extended and contemporarily recovered the outcome of GR.

\begin{figure}[htbp]
  \centering
  \includegraphics[scale=1]{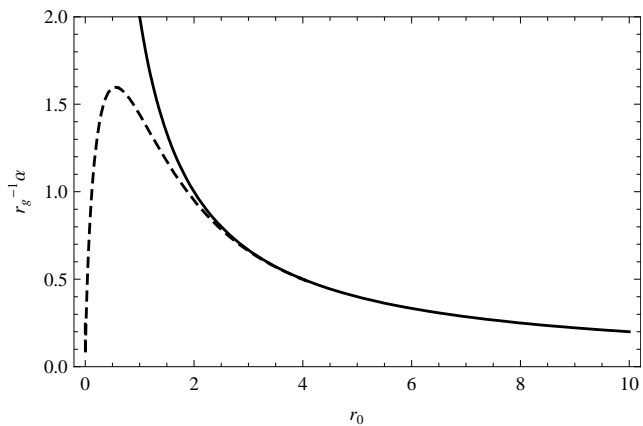}\\
  \caption{Comparison between the deflection angle of GR (solid line) and one of $f(X,Y,Z)$-Gravity (dashed line) (\ref{angle_deflec_2}) for a fixed value $\mu_2\,=\,2$ and any $\mu_1$.}
  \label{plot_1}
  \end{figure}

The analytical dependence of function $\mathcal{F}_{\mu_1,\,\mu_2}(r_0)$ from the parameters $\mu_1$ and $\mu_2$ is given by evaluating the integral (\ref{fun1}). A such as integral is not easily valuable from the analytical point of view. However this aspect is not fundamental, since we can numerically appreciate the deviation from the outcome of GR. In fact in Fig. \ref{plot_1} we show the plot of deflection angle (\ref{angle_deflec_2}) by $f(X,Y,Z)$-Gravity for a given set of values for $\mu_1$ and $\mu_2$. The spatial behavior of $\alpha$ is ever the same if we do not modify $\mu_2$. This outcome is really a surprise: by the numerical evaluation of the function $\mathcal{F}_{\mu_1,\,\mu_2}(r_0)$ one notes that the dependence of $\mu_1$ is only formal. If we solve analytically the integral we must find a $\mu_1$ independent function. However, this statement should not be justified only by numerical evaluation but it needs an analytical proof. For these reasons in the next section we reformulate the theory of GL generated by a generic matter distribution and demonstrate that for $f(X)$-Gravity one has the same outcome of GR.

\subsection{Extended Matter Source}

In this section we want to recast the framework of GL for a generic matter source distribution $\rho(\mathbf{x})$ so the photon can undergo many deviations. In this case we leave the hypothesis that the flight of photon belongs always to the same plane, but we consider only the deflection angle as the angle between the directions of incoming and leaving photon. Finally we find the generalization of GL in $f(X,Y,Z)$-Gravity including the previous outcome of deflecting point-like source (and resolving the integral (\ref{fun1})).

The relativistic invariant (\ref{me0}) is yet valid since we consider the superposition of point-like solutions. Indeed we can generalize the metric (\ref{me1}) by the following substitution

\begin{eqnarray}\label{sol_gen}
\left\{\begin{array}{ll}
\Phi\,=\,-\,G\int
d^3\mathbf{x}'\frac{\rho(\mathbf{x}')}{|\mathbf{x}-\mathbf{x}'|}
\biggl[1+\frac{1}{3}\,e^{-\mu_1|\mathbf{x}-\mathbf{x}'|}-\frac{4}{3}\,e^{-\mu_2|\mathbf{x}-\mathbf{x}'|}\biggr]
\\\\
\Psi\,=\,-\,G\int
d^3\mathbf{x}'\frac{\rho(\mathbf{x}')}{|\mathbf{x}-\mathbf{x}'|}
\biggl[1-\frac{1}{3}\,e^{-\mu_1|\mathbf{x}-\mathbf{x}'|}-\frac{2}{3}\,e^{-\mu_2|\mathbf{x}-\mathbf{x}'|}\biggr]
\end{array}\right.
\end{eqnarray}
This approach is correct only in the Newtonian limit since a such limit correspond also to the linearized version of theory. Obviously the $f(X,Y,Z)$-Gravity (like GR) is not linear, then we should have to solve the field equations (\ref{fieldequationFOG}) with a given $\rho$.

By introducing the four velocity $u^\mu\,=\,\dot{x}^\mu\,=\,(u^0, \mathbf{u})$ the flight of photon, from the metric (\ref{me0}), is regulated by the condition

\begin{eqnarray}\label{travel_photon}
g_{\alpha\beta}u^\alpha u^\beta\,=\,(1+2\Phi){u^0}^2-(1-2\Psi)|\mathbf{u}|^2\,=\,0
\end{eqnarray}
then $u^\mu$ is given by

\begin{eqnarray}\label{four_velocity}
u^\mu\,=\,\biggl(\sqrt{\frac{1-2\Psi}{1+2\Phi}}\,|\mathbf{u}|,\mathbf{u}\biggr)
\end{eqnarray}
In the Newtonian limit we find that the geodesic motion equation becomes

\begin{eqnarray}\label{geodesic_motion}
\dot{u}^\mu+\Gamma^\mu_{\alpha\beta}u^\alpha u^\beta\,=\,0\,\rightarrow\,\dot{\mathbf{u}}+|\mathbf{u}|^2\nabla(\Phi+\Psi)-2\mathbf{u}\nabla\Psi\cdot\mathbf{u}\,=\,0
\end{eqnarray}
and by supposing $|\mathbf{u}|^2\,=\,1$ we can recast the equation in a more known aspect

\begin{eqnarray}\label{geodesic_motion_1}
\dot{\mathbf{u}}\,=\,-2\biggl[\nabla_\bot\Psi+\frac{1}{2}\nabla(\Phi-\Psi)\biggr]
\end{eqnarray}
where $\nabla_\bot\,=\,\nabla-\biggl(\frac{\mathbf{u}}{|\mathbf{u}|}\cdot\nabla\biggr)\frac{\mathbf{u}}{|\mathbf{u}|}$ is the two dimensions nabla operator orthogonal to direction of vector $\mathbf{u}$. In GR we would had only $\dot{\mathbf{u}}\,=\,-2\,\nabla_\bot\Phi$ since we have $\Psi\,=\,\Phi$. In fact the field equations (\ref{NL-field-equation_2}) are corrects \cite{PRD2} if we satisfy a constraint condition among the metric potentials $\Phi$, $\Psi$ as follows

\begin{eqnarray}\label{cond}
\triangle(\Phi-\Psi)\,=\,\frac{{m_1}^2-{m_2}^2}{3{m_1}^2}\int
d^3\mathbf{x}'\mathcal{G}_2(\mathbf{x},\mathbf{x}')\,
\triangle_{\mathbf{x}'}X^{(2)}(\mathbf{x}')
\end{eqnarray}
We can affirm, then, that only in GR the metric potentials are equals (or more generally their difference must be proportional to function $|\mathbf{x}|^{-1}$). The constraint (\ref{cond}) has been found also many times in the context of cosmological
perturbation theory \cite{sugg1,sugg2,sugg3,sugg4,sugg5}.

The deflection angle (\ref{angle_deflec}) is now defined by equation

\begin{eqnarray}\label{angle_vect}
\vec{\alpha}\,=\,-\int_{\lambda_i}^{\lambda_f}\frac{d\mathbf{u}}{d\lambda}d\lambda
\end{eqnarray}
where $\lambda_i$ and $\lambda_f$ are the initial and final value of affine parameter \cite{schneider}. For a generic matter distribution we can not \emph{a priori} claim that the deflection angle belongs to lens plane (as point-like source), but we can only link the deflection angle to the difference between the initial and final velocity $\mathbf{u}$. So we only analyze the directions of photon before and after the interaction with the gravitational mass. Then the (\ref{angle_vect}) is placed by assuming $\vec{\alpha}\,=\,\Delta\mathbf{u}\,=\,\mathbf{u}_i-\mathbf{u}_f$. From the geodesic equation (\ref{geodesic_motion_1}) the deflection angle becomes

\begin{eqnarray}\label{angle_vect_1}
\vec{\alpha}\,=\,2\,\int_{\lambda_i}^{\lambda_f}\biggl[\nabla_\bot\Psi+\frac{1}{2}\nabla(\Phi-\Psi)\biggr]d\lambda
\end{eqnarray}
The formula (\ref{angle_vect_1}) represents the generalization of deflection angle in the framework of GR. By considering the photon incoming along the z-axes we can set $\mathbf{u}_i\,=\,(0,0,1)$. Moreover we decompose the general vector $\mathbf{x}\,\epsilon\,\mathbb{R}^3$ in two components: $\vec{\xi}\,\epsilon\,\mathbb{R}^2$ and $z\,\epsilon\,\mathbb{R}$. The differential operator now can be decomposed as follows $\nabla\,=\,\nabla_\bot+\hat{z}\,\partial_z\,=\,\nabla_{\vec{\xi}}+\hat{z}\,\partial_z$, while the modulus of distance is $|\mathbf{x}-\mathbf{x}'|\,=\,\sqrt{|\vec{\xi}-\vec{\xi}'|^2+(z-z')^2}\,\doteq\,\Delta(\vec{\xi},\vec{\xi}',z,z')$. Since the potentials $\Phi\,,\Psi\,\ll\,1$, around the lens, the solution of (\ref{geodesic_motion_1}) with the initial condition $\mathbf{u}_i\,=\,(0,0,1)$ can be expressed as follows

\begin{eqnarray}\label{param}
\mathbf{u}\,=\,(\mathcal{O}(\Phi,\Psi),\mathcal{O}(\Phi,\Psi),1+\mathcal{O}(\Phi,\Psi))
\end{eqnarray}
and we can substitute the integration with respect to the affine parameter $\lambda$ with $z$. In fact we note

\begin{eqnarray}\label{param1}
d\lambda\,=\,\frac{dz}{dz/d\lambda}\,=\,\frac{dz}{1+\mathcal{O}(\Phi,\Psi)}\,\sim\,dz
\end{eqnarray}
and the deflection angle (\ref{angle_vect_1}) becomes

\begin{eqnarray}\label{angle_vect_2}
\vec{\alpha}\,=\,\int_{z_i}^{z_f}\biggl[\nabla_{\vec{\xi}}(\Phi+\Psi)+\hat{z}\,\partial_z(\Phi-\Psi)\biggr]dz
\end{eqnarray}
From the expression of potentials (\ref{sol_gen}) we find the relations

\begin{eqnarray}\label{sys}
\left\{\begin{array}{ll}
\Phi+\Psi\,=\,-2\,G\int
d^2\vec{\xi}'dz'\frac{\rho(\vec{\xi}',z')}{\Delta(\vec{\xi},\vec{\xi}',z,z')}+2\,G\int
d^2\vec{\xi}'dz'\frac{\rho(\vec{\xi}',z')}{\Delta(\vec{\xi},\vec{\xi}',z,z')}\,e^{-\mu_2\Delta(\vec{\xi},\vec{\xi}',z,z')}
\\\\
\Phi-\Psi\,=\,-\,\frac{2G}{3}\int
d^2\vec{\xi}'dz'\frac{\rho(\vec{\xi}',z')}{\Delta(\vec{\xi},\vec{\xi}',z,z')}
\biggl[e^{-\mu_1\Delta(\vec{\xi},\vec{\xi}',z,z')}-e^{-\mu_2\Delta(\vec{\xi},\vec{\xi}',z,z')}\biggr]
\end{array}\right.
\end{eqnarray}
and the equation (\ref{angle_vect_1}) becomes

\begin{eqnarray}\label{angle_vect_3}
\vec{\alpha}\,&=&2G\int_{z_i}^{z_f}d^2\vec{\xi}'dz'\,dz\frac{\rho(\vec{\xi}',z')(\vec{\xi}-\vec{\xi}')}{{\Delta(\vec{\xi},\vec{\xi}',
z,z')}^3}
-2G\int_{z_i}^{z_f}d^2\vec{\xi}'dz'\,dz\frac{\rho(\vec{\xi}',z')[1+\mu_2\Delta(\vec{\xi},\vec{\xi}',z,z')]}
{{\Delta(\vec{\xi},\vec{\xi}',z,z')}^3}e^{-\mu_2\Delta(\vec{\xi},\vec{\xi}',z,z')}(\vec{\xi}-\vec{\xi}')\nonumber\\\\\nonumber
&+&\frac{2G}{3}\hat{z}\int_{z_i}^{z_f}d^2\vec{\xi}'dz'\,dz\frac{\rho(\vec{\xi}',z')(z-z')}
{{\Delta(\vec{\xi},\vec{\xi}',z,z')}^3}\biggl[\biggl(1+\mu_1\Delta(\vec{\xi},\vec{\xi}',z,z')\biggr)e^{-\mu_1\Delta(\vec{\xi},\vec{\xi}
',z,z')}
-\biggl(1+\mu_2\Delta(\vec{\xi},\vec{\xi}',z,z')\biggr)e^{-\mu_2\Delta(\vec{\xi},\vec{\xi}',z,z')}\biggr]
\end{eqnarray}
In the case of hypothesis of thin lens belonging to plane $(x,y)$ we can consider a weak dependence of modulus $\Delta(\vec{\xi},\vec{\xi}',z,z')$ into variable $z'$ so there is only a trivial error if we set $z'\,=\,0$. With this hypothesis the integral into $z'$ is incorporated by definition of two dimensional mass density $\Sigma(\vec{\xi}')\,=\,\int dz'\rho(\vec{\xi}',z')$. Since we are interesting only to the GL performed by one lens we can extend the integration range of $z$ between $(-\infty,\infty)$. Now the deflection angle is the following

\begin{eqnarray}\label{angle_vect_4}
\vec{\alpha}\,=\,4G\int d^2\vec{\xi}'\Sigma(\vec{\xi}')\biggl[\frac{1}{|\vec{\xi}-\vec{\xi}'|}-|\vec{\xi}-\vec{\xi}'|\,
\mathcal{F}_{\mu_2}(\vec{\xi},\vec{\xi}')\biggr]
\frac{\vec{\xi}-\vec{\xi}'}{|\vec{\xi}-\vec{\xi}'|}
\end{eqnarray}
where

\begin{eqnarray}\label{fun2}
 \mathcal{F}_{\mu_2}(\vec{\xi},\vec{\xi}')\,=\,\int_0^{\infty}dz\frac{(1+\mu_2\Delta(\vec{\xi},\vec{\xi}',z,0))}
 {{\Delta(\vec{\xi},\vec{\xi}',z,0)}^3}e^{-\mu_2\Delta(\vec{\xi},\vec{\xi}',z,0)}
\end{eqnarray}
The last integral in (\ref{angle_vect_3}) is vanishing because the integrating function is odd with respect to variable $z$. The expression (\ref{angle_vect_4}) is the generalization of outcome (\ref{angle_deflec_2}) and mainly we found a correction term depending only on the $\mu_2$ parameter.

In the case of point-like source $\Sigma(\vec{\xi}')\,=\,M\,\delta^{(2)}(\vec{\xi}')$ we find

\begin{eqnarray}\label{angle_vect_5}
\vec{\alpha}\,=\,2\,r_g\biggl[\frac{1}{|\vec{\xi}|}-|\vec{\xi}|\,\mathcal{F}_{\mu_2}(\vec{\xi},0)\biggr]
\frac{\vec{\xi}}{|\vec{\xi}|}
\end{eqnarray}
and in the case of $f(X,Y,Z)\,\rightarrow\,f(X)$ (\emph{i.e.} $\mu_2\,\rightarrow\,\infty$ and $\mathcal{F}_{\mu_2}(\vec{\xi},\vec{\xi}')\,\rightarrow\,0$) we recover the outcome of GR $\vec{\alpha}\,=\,2\,r_g\,\vec{\xi}/|\vec{\xi}|^2$. From the theory of GL in GR we know that the deflection angle $2\,r_g/r_0$ is formally equal to $2\,r_g/|\vec{\xi}|$ if we suppose $r_0\,=\,|\vec{\xi}|$. Besides both $r_0$, $|\vec{\xi}|$ are not practically measurable, while it is possible to measure the so-called impact parameter $b$ (see Fig. \ref{deflection}). But only in the first approximation these three quantities are equal.

In fact when the photon is far from the gravitational source we can parameterize the trajectory as follows

\begin{eqnarray}
\begin{aligned}
\begin{cases}
t\,=\,\lambda\\x\,=\,-\,t\\y\,=\,b
\end{cases}
&\rightarrow
\begin{cases}
r\,=\,\sqrt{t^2+b^2}\\\phi\,=\,-\arctan\frac{b}{t}
\end{cases}
\end{aligned}
\end{eqnarray}
and from the definition of angular momentum (\ref{motion_constant}) in the case of $t\,\gg\,b$ we have

\begin{eqnarray}
J\,=\,\dot{\phi}\,r^2\,=\,\frac{b/t^2}{1+b^2/t^2}(t^2+b^2)\,\sim\,b
\end{eqnarray}
By using the condition (\ref{Jcond}) $\dot{r}_{\pm}(r_0)\,=\,0$ we find the relation among $b$ and $r_0$\footnote{The constant $\mathcal{T}$ is dimensionless if we consider that $\lambda$ is the length of trajectory of photon. In this case without losing the generality we can choose $\mathcal{T}\,=\,1$.}

\begin{eqnarray}\label{impact_par}
b\,=\,\frac{r_0\,\mathcal{T}}{\sqrt{1-\frac{r_g}{r_0}\,\Xi(r_0)}}\,\sim\,r_0
\end{eqnarray}
justifying then the position $r_0\,=\,|\vec{\xi}|$ in the limit $r_g/r\,\ll\,1$ (but also $r_g/r_0\,\ll\,1$).

\begin{figure}[htbp]
  \centering
  \includegraphics[scale=1]{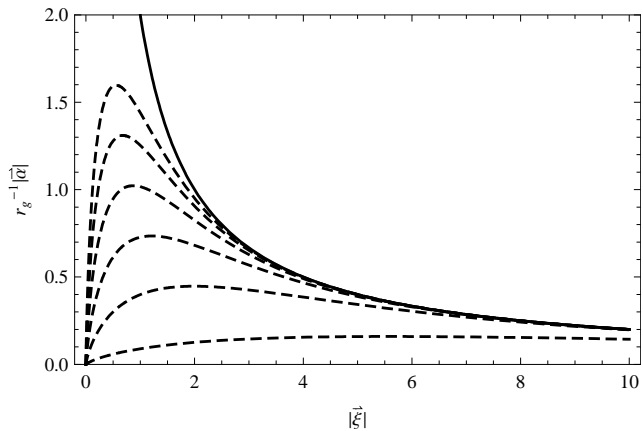}\\
  \caption{Comparison between the deflection angle of GR (solid line) and of $f(X,Y,Z)$-Gravity (dashed line) (\ref{angle_vect_5}) for $0.2\,<\,\mu_2\,<\,2$.}
  \label{plot_2}
\end{figure}

In Fig. \ref{plot_2} we report the plot of deflection angle (\ref{angle_vect_5}). The behaviors shown in figure are parameterized only by $\mu_2$ and we note an equal behavior shown in Fig. \ref{plot_1}.
With the expression (\ref{angle_vect_5}) we have the analytical proof of statement at the end of previous section. In fact in the equation (\ref{angle_vect_5}) we have not any information about the correction induced in the action (\ref{HOGaction}) by a generic function of Ricci scalar ($f_{XX}\,\neq\,0$). This result is very important if we consider only the class of theories $f(X)$-Gravity. In this case, since $\mu_2\,\rightarrow\,\infty$, we found the same outcome of GR. From the behavior in Fig. \ref{plot_2} we note that the correction to outcome of GR is deeply different for $r_0\,\rightarrow\,0$, while for $r_0\,\rightarrow\,\infty$ the behavior (\ref{angle_vect_5}) approaches one of GR, but the deviations are smaller. This difference is given by the repulsive correction to the gravitational potential (see metric (\ref{me1})) induced by $f(Y,Z)$. Only by leaving the thin lens hypothesis (the lens does not belong to plane $z\,=\,0$) we can have the deflection angle depending by $\mu_1$ (\ref{angle_vect_3}). In fact in this case the third integral in (\ref{angle_vect_3}) is not zero. Then in the case of thin lens we have a complete degeneracy of outcomes in the $f(X)$-Gravity: \emph{all $f(X)$-Gravities are equivalent to the GR}. If we want to find some differences we must to include the contributions generated by the Ricci tensor square. But also in this case we do not have the right behavior: the deflection angle is smaller than one of GR: $f(X,Y,Z)$-Gravity does not mimic the Dark Matter component if we assume the thin lens hypothesis.

\subsection{Lens equation}

\begin{figure}[htbp]
  \centering
  \includegraphics[scale=1]{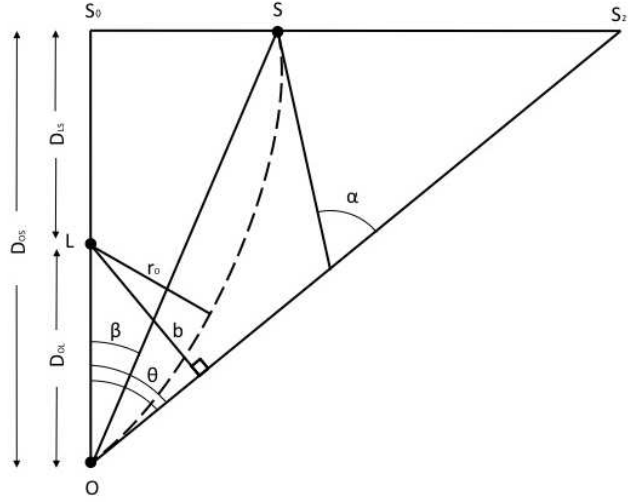}\\
  \caption{The gravitational lensing geometric for a point-like source lens $L$ at distance $D_{OL}$ from observer $O$. A source $S$ at distance $D_{OS}$ from $O$ has angular position $\beta$ from the lens. A light ray (dashed line) from $S$ which passes the lens at minimal distance $r_0$ is deflected by $\alpha$; the observer sees an image of the source at angular position $\theta\,=\,b/D_{OL}$ where $b$ is the impact factor. $D_{LS}$ is the distance lens - source.}
  \label{deflection}
\end{figure}

To demonstrate the effect of a deflecting mass we show in Fig. \ref{deflection} the simplest GL configuration. A point-like mass is located at distance $D_{OL}$ from the observer $O$. The source is at distance $D_{OS}$ from the observer, and its true angular separation from the lens $L$ is $\beta$, the separation which would be observed in the absence of lensing ($r_g\,=\,0$). The photon which passes the lens at distance $r_0\,\sim\,b$ is deflected with an angle $\alpha$.

Since the deflection angle (\ref{angle_deflec_2}) is equal to (\ref{angle_vect_5}), for sake of simplicity we will use the "vectorial" expression. Then the expression (\ref{angle_vect_5}), by considering the relation (\ref{impact_par}), becomes

\begin{eqnarray}\label{angle_vect_6}
\alpha\,=\,2\,r_g\biggl[\frac{1}{b}-b\,\mathcal{F}_{\mu_2}(b,0)\biggr]
\end{eqnarray}
The condition that this photon reach the observer is obtained from the geometry of Fig. \ref{deflection}. In fact we find
\begin{eqnarray}\label{lens_eq_1}
\beta\,=\,\theta-\frac{D_{LS}}{D_{OS}}\,\alpha
\end{eqnarray}
Here $D_{LS}$ is the distance of the source from the lens. In the simple case with a Euclidean background metric here, $D_{LS}\,=\,D_{OS}-D_{OL}$; however, since the GL occurs in the Universe on large scale, one must use a cosmological model \cite{schneider}. Denoting the angular separation between the deflecting mass and the deflected photon as $\theta\,=\,b/D_{OL}$ the lens equation for $f(X,Y,Z,)$-Gravity is the following

\begin{eqnarray}\label{lens_eq_2}
[1+{\theta_E}^2\,\mathcal{F}(\theta)]\,\theta^2-\beta\,\theta-{\theta_E}^2\,=\,0
\end{eqnarray}
where $\theta_E\,=\,\sqrt{\frac{2\,r_g\,D_{LS}}{D_{OL}D_{OS}}}$ is the Einstein angle and

\begin{eqnarray}\label{fun3}
\mathcal{F}(\theta)\,=\,\int_0^{\infty}dz\frac{(1+\mu_2 D_{OL}\sqrt{\theta^2+z^2})}{\sqrt{(\theta^2+z^2)^3}}\,e^{-\mu_2 D_{OL}\sqrt{\theta^2+z^2})}
\end{eqnarray}

\begin{figure}[htbp]
  \centering
  \includegraphics[scale=1]{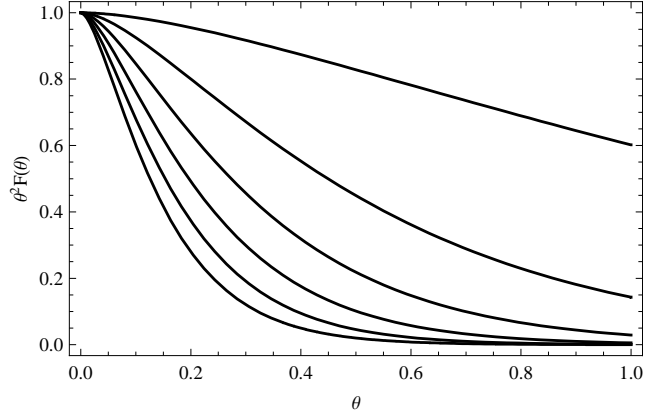}\\
  \caption{Plot of function $\theta^2\mathcal{F}(\theta)$ (\ref{fun3}) for $1\,<\,\mu_2D_{OL}\,<\,10$.}
  \label{plot_3}
\end{figure}

\begin{figure}[htbp]
  \centering
  \includegraphics[scale=1]{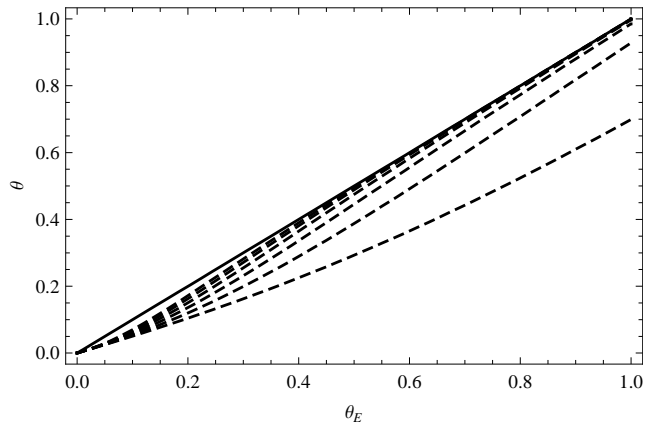}\\
  \caption{Plot of the Einstein ring (solid line) and its modification (\ref{lens_sol_ring}) in the $f(X,Y,Z)$-Gravity for $1\,<\,\mu_2D_{OL}\,<\,10$ (dashed line).}
  \label{plot_4}
\end{figure}

Since we have $0\,<\,\theta^2\mathcal{F}(\theta)\,<\,1$ (Fig. \ref{plot_3}) we can find a perturbative solution of (\ref{lens_eq_2}) by starting from one in GR, $\theta^{GR}_\pm\,=\,\frac{-\beta\pm\sqrt{\beta^2+4{\theta_E}^2}}{2}$. In fact by assuming $\theta\,=\,\theta^{GR}_\pm+\theta^*$ and neglecting ${\theta^*}^2\mathcal{F}(\theta^*)$ in (\ref{lens_eq_2}) we find

\begin{eqnarray}\label{lens_sol}
\theta\,=\,\theta^{GR}_\pm\mp\frac{{\theta_E}^2}{\sqrt{\beta^2+4\,{\theta_E}^2}}\,\mathcal{F}(\theta^{GR}_\pm)\,
{\theta^{GR}_\pm}^2
\end{eqnarray}
and in the case of $\beta\,=\,0$ we find the modification to the Einstein ring

\begin{eqnarray}\label{lens_sol_ring}
\theta\,=\,\pm\theta_E\biggl[1-\frac{{\theta_E}^2}{2}\,\mathcal{F}(\theta_E)\biggr]
\end{eqnarray}
In Fig. \ref{plot_4} we show the angular position of images with respect to the Einstein ring. Both the deflection angle and the position of images assume a smaller value than ones of GR. Then the corrections to the GR quantities are found only for the introduction in the action (\ref{HOGaction}) of curvature invariants $Y$ (or $Z$), while there are no modifications induced by adding a generic function of Ricci scalar $X$. The algebraic signs of terms concerning the parameter $\mu_2$ are ever different with respect to the terms of GR in (\ref{me1}) and they can be interpreted as a "repulsive force" giving us a minor curvature of photon. The correction terms concerning the parameter $\mu_1$ have opposite algebraic sign in the metric component $g_{tt}$ and $g_{ij}$ (\ref{me1}) and we lose their information in the deflection angle (\ref{angle_vect_2}).


In both approaches we find the same outcomes $\mu_1$-independent because the matter source (in our case it is a point-like mass) is symmetric with respect to $z$-axes and we neglect the second integral in (\ref{angle_vect_3}). Obviously for a generic matter distribution the deflection angle is defined by (\ref{angle_vect_3}) and the choice of second derivative of function of Ricci scalar is not arbitrary anymore.

\section{Conclusions}

In this paper we compute the study of GL when a FOG is considered. Among the several theories of fourth order we consider a generic function of Ricci scalar, Ricci and Riemann tensor, but by using Gauss-Bonnet invariant it is adequate to consider only a $f(X,Y)$-Gravity.

We start from the outcome of previous paper \cite{PRD2} about the point-like solutions in the so-called Newtonian limit of theory and formulate the deflection angle and the angular positions of image. The study has been evaluated on two steps: in the first one we consider a point-like source and by analyzing the properties of Lagrangian of photon we obtain a correction to the outcome of GR depending apparently on two free parameters of theory. But by plotting, only numerically, the new angular behavior (\ref{angle_deflec_2}) with respect to the minimal distance $r_0$ we note that the correction term does not depend on the parameter $\mu_1$ (Fig. \ref{plot_1}). In the second step we start by more general geodesic motion and we reformulate the deflection angle for a generic matter distribution. In the case of an axially symmetric matter density we obtain the usual relation between the deflection angle and the orthogonal gradient of metric potentials (\ref{angle_vect_4}). Otherwise we find that the angle is depending also onto the travel direction of photon (\ref{angle_vect_3}). Particularly if there is a $z$-symmetry the deflection angle does not depend explicitly on the parameter $\mu_1$ but we have only the correction term induced by $\mu_2$ (\ref{fun2}).

From the definition of $\mu_1$ and $\mu_2$ (\ref{mass_definition}) we note that the presence of function of Ricci scalar ($f_{XX}(0)\,\neq\,0$) is only in $\mu_1$. \emph{Then if we consider only the $f(X)$-Gravity ($\mu_2\,\rightarrow\,\infty$) the geodesic trajectory of photon is unaffected by the modification in the Hilbert-Einstein action}. Our analysis is compatible with respect to the principal outcome shown in the paper \cite{JET}. Instead if we want to have the corrections to GR it needs to add a generic function of Ricci tensor square into Hilbert-Einstein action. But in this case we find the deflection angle smaller than one of GR (\ref{angle_deflec_2}) or (\ref{angle_vect_5}). Obviously the same situation is present also in the Einstein ring (\ref{lens_sol_ring}), where the new angle is ever lower than the one of GR (Fig. \ref{plot_4}). The mathematical motivation is a consequence of algebraic signs of terms containing the parameter $\mu_2$ in the metric (\ref{me1}). In fact they are ever different with respect to the terms of GR in (\ref{me1}) and they can be interpreted as a "repulsive force" giving us a minor curvature of photon trajectory, instead the correction terms containing the parameter $\mu_1$ have opposite algebraic sign in the metric components $g_{tt}$ and $g_{ij}$ (\ref{me1}) and we lose their information in the deflection angle (\ref{angle_vect_2}).

A similar outcome has been found for the galactic rotation curve \cite{Stabile_scelza}, where the contribution of $f(Y)$ in the action gives us a lower rotation velocity profile than the one of GR, but with a no trivial difference. In fact in galactic dynamics we are studying the motion of massive particles and in this case we find the corrections induced also by $f(X)$-Gravity. Then if we can estimate the weight of the corrections (induced by $f(X)$-Gravity) to the Ricci scalar for the galactic motion, from the point of view of GL we have a perfect agreement with the GR. Only by adding $f(Y)$ in the action we induce the modifications in both two frameworks, but we do not find the hoped behavior: the flat galactic rotation curve and a more strong deflection angle of photon. Also for a photon bending we need a Dark Matter component. \emph{Moreover if we consider $f(X,Y)$-Gravity for the GL we need a bigger amount of Dark Matter then in GR}. A such conclusion qualitatively does not differ by the one about the galactic rotation curves \cite{Stabile_scelza}.

Also after the analysis of GL we can affirm, as for the galactic rotation curves, it remains a hard challenge to interpret the Dark Matter effects as a single geometric phenomenon.

\section{Acknowledges}

We would like to dedicate this first our paper to our dear parents.

\end{document}